\documentclass[journal]{IEEEtran}
\usepackage[pdftex]{graphicx}
\usepackage{threeparttable}
\usepackage{array}
\usepackage{rotating}
\usepackage{multirow}
\usepackage{afterpage}
\usepackage{color,soul}
\usepackage{fixltx2e}
\usepackage[cmex10]{amsmath}
\DeclareMathOperator*{\argmin}{\arg\!\min}
\usepackage{mathtools}
\usepackage{url}
\begin{document}

\title{Performance Evaluation of Frequency Domain Equalisation Based Colour Shift Keying Modulation Schemes Over Diffuse Optical Wireless Channels}

\author{R.~Singh, \emph{Member, IEEE}, T.~O'Farrell, \emph{Senior Member, IEEE}, T.~C. Bui, \emph{Student Member, IEEE},\\ M.~Biagi, \emph{Senior Member, IEEE}, and~J.~P.~R.~David, \emph{Fellow, IEEE}
\thanks{R. Singh is with the Department of Engineering Science, University of Oxford, Oxford, OX1 3PJ, UK. (e-mail: \emph{ravinder.singh@eng.ox.ac.uk}). This work was carried out during his PhD at the University of Sheffield.}
\thanks{T.C. Bui and M. Biagi are with the Department of Information, Electrical and Telecommunication Engineering, Sapienza University of Rome, Italy.}
\thanks{T. O'Farrell and J. P. R. David are with the Department of Electronic and Electrical Engineering, University of Sheffield, Sheffield, S1 3JD, UK.}}

\maketitle
\begin{abstract}
Multi-colour light-emitting diode (LED) based visible light communication (VLC) benefits from wavelength diversity while providing indoor illumination. Colour shift keying (CSK) is a well-researched IEEE standardised multi-colour VLC modulation technique. This paper presents an investigation into the performance of tri-chromatic LED (TLED) CSK standardised in IEEE 802.15.7 and a quad-chromatic LED (QLED) CSK, over a range of diffuse optical wireless channels, and proposes the use of frequency domain equalisation (FDE) at the receiver to combat multipath dispersion. The investigation results show that the FDE enables the higher order CSK modulation modes to achieve a bit error rate (BER) of 10\textsuperscript{-6} for finite amounts of optical power while operating over highly dispersive channels and hence provide data links of up to 85.33 and 256 Mbit/s through TLED and QLED CSK, respectively, for a system bandwidth of 24 MHz. The overall optical power requirements of the CSK schemes can be reduce by up to 12.6 dB with the use of FDE at the cost of a small overhead due to cyclic prefix. The optical channel model used in this investigation includes the cross-talk and insertion losses caused by the optical properties of commercially available system front end devices.  

\end{abstract}

\section{Introduction}
\IEEEPARstart{T}{he energy} efficiency and longer life-spans has set the LEDs to dominate as the lighting source around the world. At the same time, the advancements in the modulation techniques and optoelectronic devices for VLC are promising a bright future for the technology and provide potential solutions to the ever growing data-rate demands of wired and wireless devices \cite{rajbhandari2017review}\cite{zafar2017laser}. The use of multi-colour or multi-primary LEDs has gained a lot of interest due to their ability to achieve higher data-rates by utilising multiple visible spectrum channels. The conventional TLED CSK multi-primary modulation scheme, standardised in the IEEE 802.15.7, can provide up to 96 Mbit/s for a 24 MHz system bandwidth \cite{IEEESTD}\cite{PHYSummary} and the standard is currently under revision such that the performance of the VLC physical layer (PHY) can be improved.

Extensive performance evaluation of the standardised TLED system has been carried out \cite{ourPaper}\cite{sarbazi2013phy}\cite{singh2015higher} and research on TLED CSK constellation design optimisation has also been presented \cite{Drost}\cite{EricSteve} to improve the output colour quality. The idea of using more than three primary light sources has also been introduced \cite{MM}, and in particular, a four primary LED based QLED CSK modulation scheme \cite{JLT} has been designed to enhance the minimum Euclidean distances between the CSK symbols such that a better system BER performance can be achieved by efficiently using the available chromatic (or colour) and the intensity (or signal) spaces \cite{CIM}.

Recently, we have studied the performance of CSK systems in a line-of-sight (LOS) indoor environment using forward error correction and FDE \cite{singh2015physical}\cite{singh2018coded}\cite{singh2015analysis}. It is well known that under most operating environments, VLC can benefit from LOS channels. However, at the same time, there are many usage scenarios, where the LOS can be blocked due to indoor objects and/or a person. In such a case, the CSK schemes must provide data links through the diffuse signals detected by the photo-detector(s) (PDs) at the receiver. Until now, no work has been carried out to evaluate and analyse the performance of the CSK modulation schemes under the non line-of-sight (NLOS) channels.

Extending our work in \cite{singh2018coded}, this paper presents investigations into the error performances of both the TLED and QLED CSK systems over the diffuse wireless optical channels, both with and without the consideration of FDE. A generic model of the diffuse indoor optical wireless channel is used which provides uniform distribution of the optical power across the indoor environment, such as an office, and allows wide range of selection for the induced temporal dispersion levels. The investigation includes the cross-talk between the multi-colour channels and insertion losses incurred due to the optical properties of LEDs, receive colour filters and PDs. The performance analysis of the unequalised CSK schemes suggests that the higher order modulation modes of both CSK systems will require infinite amount of optical power to operate over channels with moderate levels of dispersion, making it impossible for the CSK to provide a high data rate link.

FDE is well known for providing a low complexity means to combat temporal dispersion, of a single carrier (SC) modulation based data signal, caused by multipath dispersion \cite{FDE1}. As in the optical OFDM \cite{oOFDM}, the FDE based CSK schemes will have high efficiency and low computational complexity. In addition, the FDE based CSK will have the advantage of lower peak to average power ratios (PAPR) at the transmitter, which will reduce the signal conditioning requirements caused by the non-linear I-V characteristics of LEDs, such as pre-distortion and power back-off \cite{Pridist1}\cite{Pridist2}.

The result analysis reveals that the FDE enables the TLED and QLED schemes to successfully operate at 85.33 Mbit/s and 256 Mbit/s over highly dispersive non-LOS diffuse optical channels, with a small amount of overhead due to the cyclic prefix. The results also show that the optical SNR requirements of the CSK schemes decrease by amounts between 0.8 to 12.6 dB through the use of FDE for a significantly low bit error rate (BER).

The rest of this paper is organised in the following manner. Section~\ref{sec:SystemDescription} describes the principle of FDE based TLED and QLED CSK modulation schemes. The details of the optical channel cross-talk are also given in Section~\ref{sec:SystemDescription}. Section~\ref{sec:inDis} details the optical wireless channel model used in this work. Section~\ref{sec:inDis} also presents and analyses the performance results of the FDE based TLED and QLED systems. Finally, section \ref{sec:conclusion} presents the conclusions.
\section{FDE and FEC based CSK Systems}
\label{sec:SystemDescription}
\subsection{CSK Modulation}
\label{sec:CSKmod}
The CSK modulation schemes are based on the \emph{x-y} colour coordinates, defined by the international commission on illumination in CIE 1931 colour space \cite{CIE}. The CIE 1931 colour space represents all the colours visible to the human eye with their chromaticity values \emph{x} and \emph{y}. In CSK, the intensity of multicolour LEDs is modulated for data transmission. The mixture of light produced from the LED sources allows CSK to regenerate various colours, each of which can be represented by a pair of \emph{x-y} coordinates and these chromaticity pairs represent different data symbols. 

For three colour LED systems, such as TLED CSK, to generate colour of each chromatic pair, the intensities required for each LED is calculated based on the linear transformation given by \cite{IEEESTD}: 
\begin{eqnarray}
\label{eq:Conversion}
\left[ \begin{array}{c} x \\ y \\ 1 \end{array} \right] =  \left[ \begin{array}{ccc} x_i & x_j & x_k \\ y_i & y_j & y_k \\ 1 & 1 & 1 \end{array} \right]\left[ \begin{array}{c} I_i \\ I_j \\ I_k \end{array} \right]
\end{eqnarray}
In equation~(\ref{eq:Conversion}), the coordinates $(x_{i}, y_{i})$, $(x_{j}, y_{j})$ and $(x_{k}, y_{k})$ refer to the CWCV of the light sources used in the system. The $(x_{i}, y_{i})$, $(x_{j}, y_{j})$ and $(x_{k}, y_{k})$ also represent one CSK symbol each, with remaining symbols each denoted by a pair of $x$ and $y$ chromaticities. The intensities for each LED are represented by $I_i$, $I_j$ and $I_k$.

\begin{figure}[tbph!]
\centering
\includegraphics[height=2in,width=0.6\linewidth]{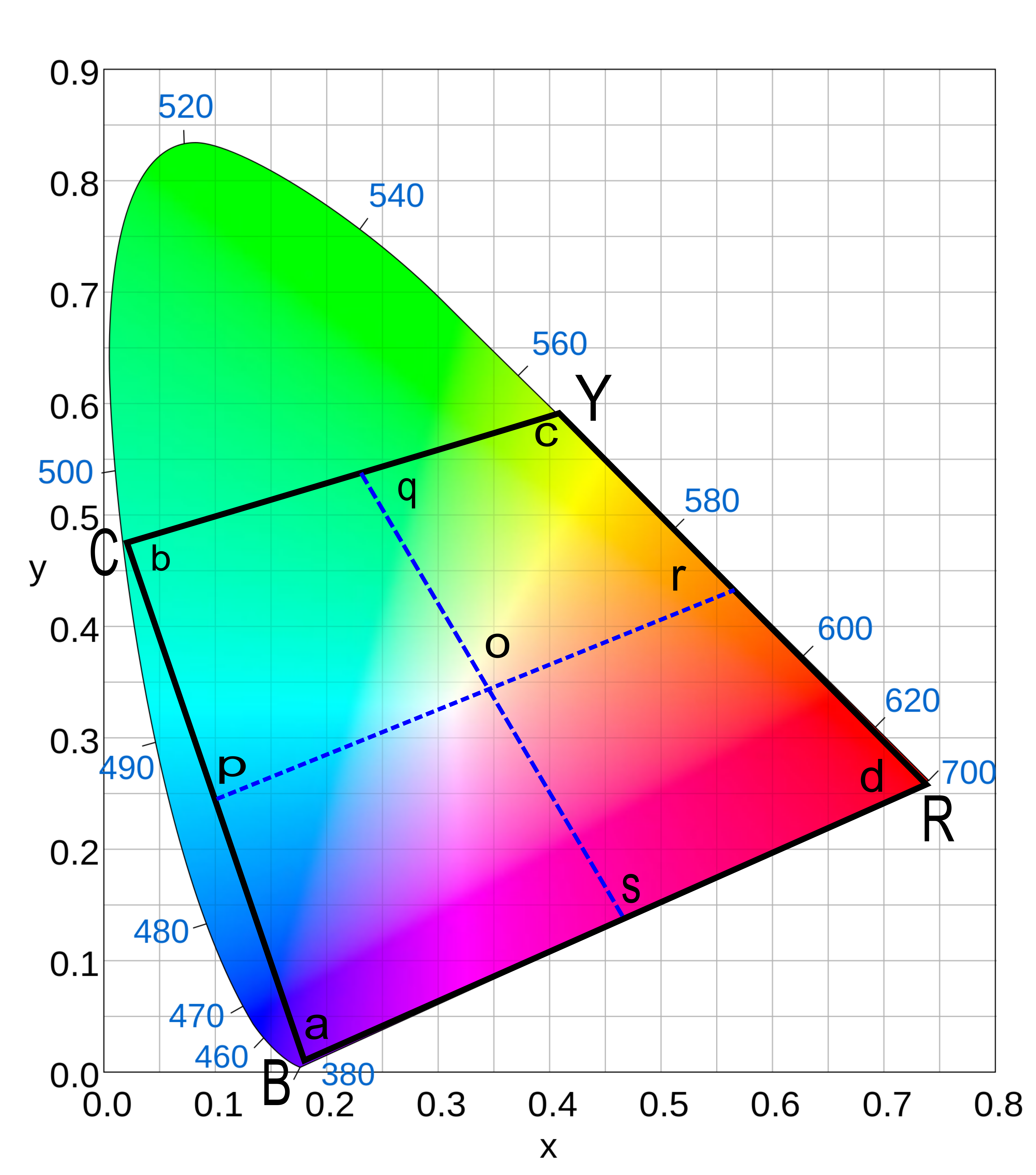}
\caption{Operational chromatic space of the QLED CSK system on the CIE 1931 x-y colour co-ordinate diagram}
\label{fig:CIErect}
\end{figure}

In QLED CSK, as four different colour LED sources are used, the symbol constellation forms a quadrilateral shape, as shown in Fig.~\ref{fig:CIErect}, where BCYR represent the central wavelength chromaticity values (CWCV) of the used blue, cyan, yellow and red LED sources. This quadrilateral constellation allows simple symbol mapping as in a conventional M-QAM scheme \cite{JLT}. On the other hand, in TLED CSK the constellation is of a triangular shape.

The QLED system is designed in a way that it uses up to three out of four LEDs for each CSK symbol. Therefore, the linear transformation of equation~(\ref{eq:Conversion}) can be used for the conversion between chromaticity to intensity \cite{JLT}. The operational space in QLED CSK, shown in Fig.~\ref{fig:CIErect}, is divided into four quadrilaterals `pbqo', `oqcr', `sord' and `apos'. The colours within each of these quadrilaterals are regenerated by mixing the light from BCY, CYR, YRB and RBC LEDs, respectively. Therefore, as in the TLED CSK, the total optical power transmitted at any instance is constant and nominally equal to one watt in the QLED CSK.  
\subsection{System Description}
\label{sec:FDEcsk}
\begin{figure*}[tbph!]
\centering
\includegraphics[width=\linewidth]{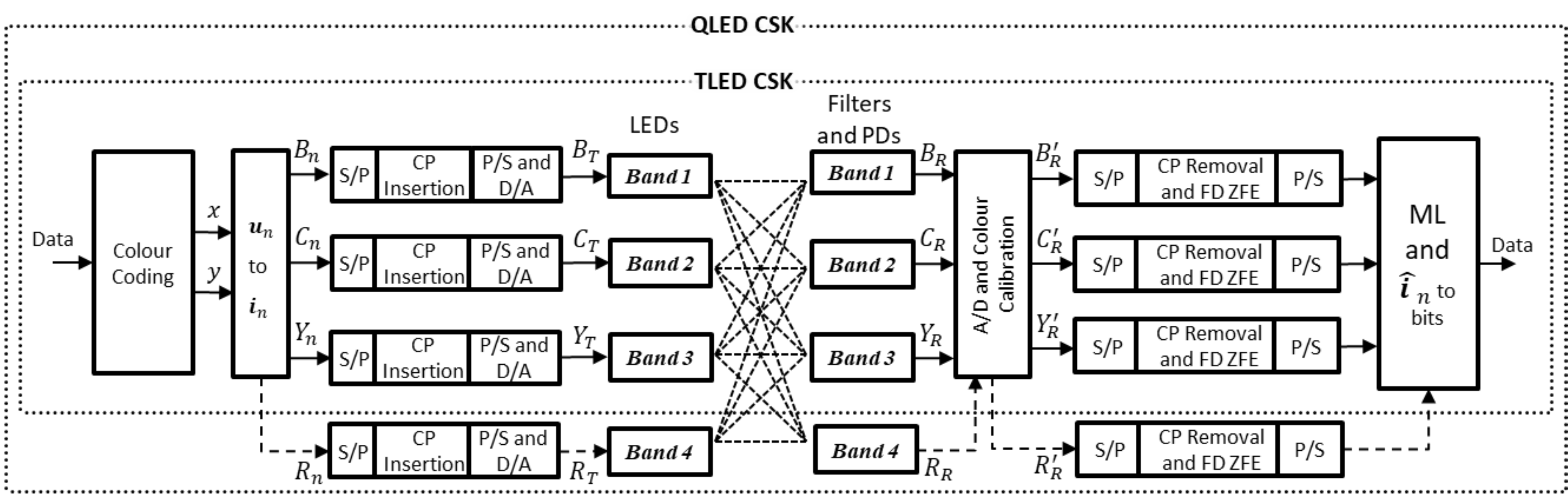}
\caption{Coded TLED and QLED CSK systems with FDE.}
\label{fig:FDEqledSys}
\end{figure*}

Fig.~\ref{fig:FDEqledSys} shows the block diagram of FDE based QLED and TLED schemes. At TX side the binary data is grouped into $k=log_2(M)$ bits, where $M$ is the modulation order, and mapped to dedicated chromaticity pairs, given by vectors ${\boldsymbol{u}}_n = [x_{n},y_{n},1]^T$, where $n$ represents the n\textsuperscript{th} chromaticity pair and $T$ represents the transpose. Based on ${\boldsymbol{u}}_n$, the n\textsuperscript{th} intensity vector ${\boldsymbol{i}}_n = [B_n,C_n,Y_n,R_n]^T$ is obtained for the four LED sources as detailed in sec.~\ref{sec:CSKmod}. The $B_n$, $C_n$, $Y_n$ and $R_n$ represent the intensity of each LED source for the n\textsuperscript{th} chromaticity based CSK symbol.

In order to enable FDE, CSK systems use block wise transmission. Therefore, each of the serial intensity values, $(B_n,C_n,Y_n,R_n)$, are parsed into vectors of length $N$, e.g. for $B_n$, a vector denoted by ${\boldsymbol{b_T}}=[B_0,B_1,B_2,....B_{N-1}]^T$. Here, $(.)_{\boldsymbol{T}}$ signifies the transmit side of the system. Similarly, the transmit intensity vectors ${\boldsymbol{c_T}}$, ${\boldsymbol{y_T}}$ and ${\boldsymbol{r_T}}$ are obtained.  Then a cyclic prefix (CP) of length $L$ is a added to the front of each of these vectors, which, for example, gives the transmit vector for the blue band as $[B_{N-L},B_{N-(L-1)},....B_{N-1},B_0,B_1,B_2,....B_{N-1}]^T$. Therefore, each block contains $N_{SB}$ number of sub-blocks, where $N_{SB}=N+L$. The parallel to serial conversion then takes place and through the use of digital to analogue converters the transmit signals are obtained which modulate the intensity of each of the LEDs source. The transmit signals are given as $B_T$, $C_T$, $Y_T$ and $R_T$. 

At the receiving end the narrowband optical filters pass light of the desired wavelength to the PDs. The received signals at the output of the PDs can be given as\footnote{The system description in this paper is presented only for the QLED scheme for brevity. The FDE based TLED system can also be described similarly, where only three LEDs and three PDs are used.}:
\setlength{\arraycolsep}{0.06cm}
\begin{eqnarray}
\label{eq:QLEDsysEq}
\left[ \begin{array}{c} B_R(t) \\ C_R(t) \\ Y_R(t) \\ R_R(t) \end{array} \right]\!\!=\!\!\overbrace{\left[\begin{array}{cccc} g_{1,1} & g_{1,2} & g_{1,3} & g_{1,4} \\ g_{2,1} & g_{2,2} & g_{2,3} & g_{2,4} \\ g_{3,1} & g_{3,2} & g_{3,3} & g_{3,4} \\ g_{4,1} & g_{4,2} & g_{4,3} & g_{4,4} \end{array}\right]}^{\boldsymbol{G}}\!\!\left[\begin{array}{c} h(t)\!*\!B_T(t) \\ h(t)\!*\!C_T(t) \\ h(t)\!*\!Y_T(t) \\ h(t)\!*\!R_T(t) \end{array}\right] \!\!+\!\! \left[\begin{array}{c} n_B(t) \\ n_C(t) \\ n_Y(t) \\ n_R(t) \end{array}\right]  
\end{eqnarray}
\setlength{\arraycolsep}{5pt}

In equation~(\ref{eq:QLEDsysEq}), `$*$' is the convolution operator, $h(t)$ is the impulse response of the VLC channel, which is further detailed in sec.~\ref{sec:Channel}. The independent identically distributed AWGN per detector, given by $[n_B,n_C,n_Y,n_R]^T$ for QLED CSK, are modelled as in \cite{JLT} and each has a noise variance of ${\sigma^2}$, where $\sigma$ is the standard deviation of noise, which is related to the single sided noise power spectral density, $N_o$ as $\sigma=\sqrt{N_o/2}$. $\boldsymbol{G}$ is a square channel cross-talk and insertion loss (CIL) matrix, where $g_{i,j}$ represents the effective responsivity between the receive band $i$ and transmit band $j$, which is calculated as:
\vspace{-0.05in} 
\begin{eqnarray}
\label{eq:EffectiveR}
g_{i,j} = {\int\limits_{\lambda_{min}^{T_i}}^{\lambda_{max}^{T_i}} S_j(\lambda)T_i(\psi,\lambda)\Re(\lambda)d\lambda} \Bigg/ {\int\limits_{\lambda_{min}^{S_j}}^{\lambda_{max}^{S_j}} S_j(\lambda)d\lambda},
\end{eqnarray}  
where, $S(\lambda)$ is the relative spectral power distribution (SPD) of the LEDs, $T(\psi,\lambda)$ is the filter transmissivity, and $\Re(\lambda)$ is the responsivity of the photo-detector(s) (PDs). 

Post analogue to digital conversion, the colour calibration as suggested in the standard \cite{IEEESTD} takes place to compensate for the cross-talk between the multi colour channels. The instantaneous sets of intensities after calibration can be given as, $[B_R^{'},C_R^{'},Y_R^{'},R_R^{'}]^T = \boldsymbol{G}^{-1}[B_R,C_R,Y_R,R_R]^T$. The received intensity signals ${B_r^{'}}$, $C_r^{'}$, $Y_r^{'}$ and $R_r^{'}$ are converted from serial to parallel for CP removal. Then each of the received intensity blocks ${\boldsymbol{b_R}}$, ${\boldsymbol{c_R}}$, ${\boldsymbol{y_R}}$ and ${\boldsymbol{r_R}}$ is converted to frequency domain using fast Fourier transform (FFT) of size $N$, so that the FDE can take place. The transmit ${\boldsymbol{b_T}}$ and receive ${\boldsymbol{b_R}}$ vectors can be mathematically related as:
\begin{equation}
\label{eq:ChanB}
{\boldsymbol{b_R}} = {\boldsymbol{H}}{\boldsymbol{b_T}}+{\boldsymbol{n_B}}
\end{equation}

In equation~{\ref{eq:ChanB}}, ${\bf{H}}$ is the $N \times N$ channel circulant convolutional matrix, which can be diagonalise as ${\boldsymbol{H}}={\boldsymbol{F}}^H {\boldsymbol{\Lambda F}}$ \cite{ZFE}, where ${\boldsymbol{F}}$ is the FFT matrix and ${\boldsymbol{\Lambda}}$ is a diagonal matrix with diagonal entries equal to the FFT of the optical channel impulse response $h(t)$ (See sec.~\ref{sec:Channel} for $h(t)$). Here, $(.)^H$ represents the Hermitian transpose. The AWGN noise vector for the blue band is given as ${\boldsymbol{n_B}}=[{n_B}_1,{n_B}_2,...{n_B}_{N-1}]^T$.

In this paper, the FDE is based on zero-forcing equaliser, ${\boldsymbol{Z}}$, which is the frequency domain equaliser matrix given as, ${\bf{Z}}= {\boldsymbol{\Lambda}}^H({\boldsymbol{\Lambda}}{\boldsymbol{\Lambda}}^H)^{-1}$. The equalised vector ${\boldsymbol{b}}$ is obtained as, ${\boldsymbol{b}} = {\boldsymbol{F}}^H{\boldsymbol{ZFb_R}}$. Similarly, the vectors  ${\boldsymbol{c}}$, ${\boldsymbol{y}}$ and ${\boldsymbol{r}}$ are obtained as well and then the parallel to serial conversion gives the \emph{n\textsuperscript{th}} received intensity vector, ${\boldsymbol{\tilde{i}}}_n = [\tilde{B_n},\tilde{C_n},\tilde{Y_n},\tilde{R_n}]^T$. At this point, the final intensity vector $\hat{\boldsymbol{i}}_n$ is obtained from each ${\boldsymbol{\tilde{i}}}_n$ through the use of maximum likelihood (ML) detection
\begin{equation}
\label{eq:HD}
\hat{\boldsymbol{i}}_n = \argmin_{\substack{\boldsymbol{i} \in \mathcal{I}}} ||\tilde{\boldsymbol{i}}_n-\boldsymbol{i}||^2,
\end{equation}
where, $\mathcal{I}$ contains the intensity based alphabets of the CSK constellation. The data bits are then de-mapped from $\hat{\boldsymbol{i}}$ to retrieve the final information.

\section{Performance Evaluation of FDE based TLED and QLED systems over dispersive optical wireless channel with AWGN}
\label{sec:inDis}
In this section, the performance of uncoded TLED and QLED CSK systems is investigated over a non-LOS dispersive optical wireless channel, with and without the use of FDE. The non-LOS case is used as a worst case scenario where the transmitted visible light signals reflect off multiple room objects and walls, and propagate through various paths with different path lengths towards the receiver. Therefore, only diffuse light signals are present at the receiver. Multiple copies of each transmitted pulse are received through the PDs at different times. The amplitude of these received copies reduce exponentially with time due to the increase in the number of reflections that each path contains. This multipath behaviour of an indoor optical wireless channel causes temporal dispersion of a transmitted optical pulse and ISI between multiple transmitted data symbols.

\subsection{Optical Channel Model}
\label{sec:Channel}
The indoor dispersive optical channel is generally modelled as an impulse response of a first order low-pass filter \cite{WIC},\cite{MoIRWC}. In this paper, the exponential-decay model has been used to represent the indoor wireless visible light channel, whose impulse response $h(t)$ can  be given as \cite{MoIRWC}:
\begin{equation}
\label{eq:Impulse}
h(t) = \frac{1}{\tau}\textmd{exp}(\frac{-t}{\tau})u(t)
\end{equation}
Where $u(t)$ is the unit step function and $\tau$ is the exponential decay time constant, which is related to the channel \emph{rms} delay spread $D_{rms}$ as $D_{rms}=\tau/2$.

For CSK systems, the properties of the optical front ends, such as the spectral response of LEDs, the optical gain and response of receive filters, and the responsivity of PDs also affect the overall system performance. In \cite{JLT}, commercially available optical front-end components were used to estimate the matrix $G$ for both the TLED and QLED schemes. In this paper, use of same optical front end systems has been assumed. Therefore, for TLED scheme: 
\begin{equation}
\label{eq:GTLED}
G = \left[ \begin{array}{ccc} 0.271 & 0.030 & 0 \\ 0 & 0.255 & 0 \\ 0 & 0 & 0.200 \end{array} \right]
\end{equation}
and for the QLED scheme:
\begin{equation}
\label{eq:GQLED}
G = \left[ \begin{array}{cccc} 0.200 & 0.003 & 0 & 0 \\ 0.007 & 0.220 & 0.003 & 0 \\ 0 & 0.002 & 0.255 & 0 \\ 0 & 0 & 0.030 & 0.271 \end{array} \right]
\end{equation}
The matrices (\ref{eq:GTLED}) and (\ref{eq:GQLED}) suggests that the insertion losses will be very high, whereas, the cross-talk will be very small for the CSK systems with the considered front-end devices. However, such insertion losses are typical of most VLC systems and largely attributable to the responsivity of the PDs. As mentioned earlier, to minimise the effect of cross-talk on the performance of the TLED and QLED systems, at the receiver, colour calibration \cite{IEEESTD} was used during the simulations. 


\subsection{Simulations and Results}
\label{sec:SimResults}
\begin{figure}[tbph!]
	\centering
	\includegraphics[width=\linewidth]{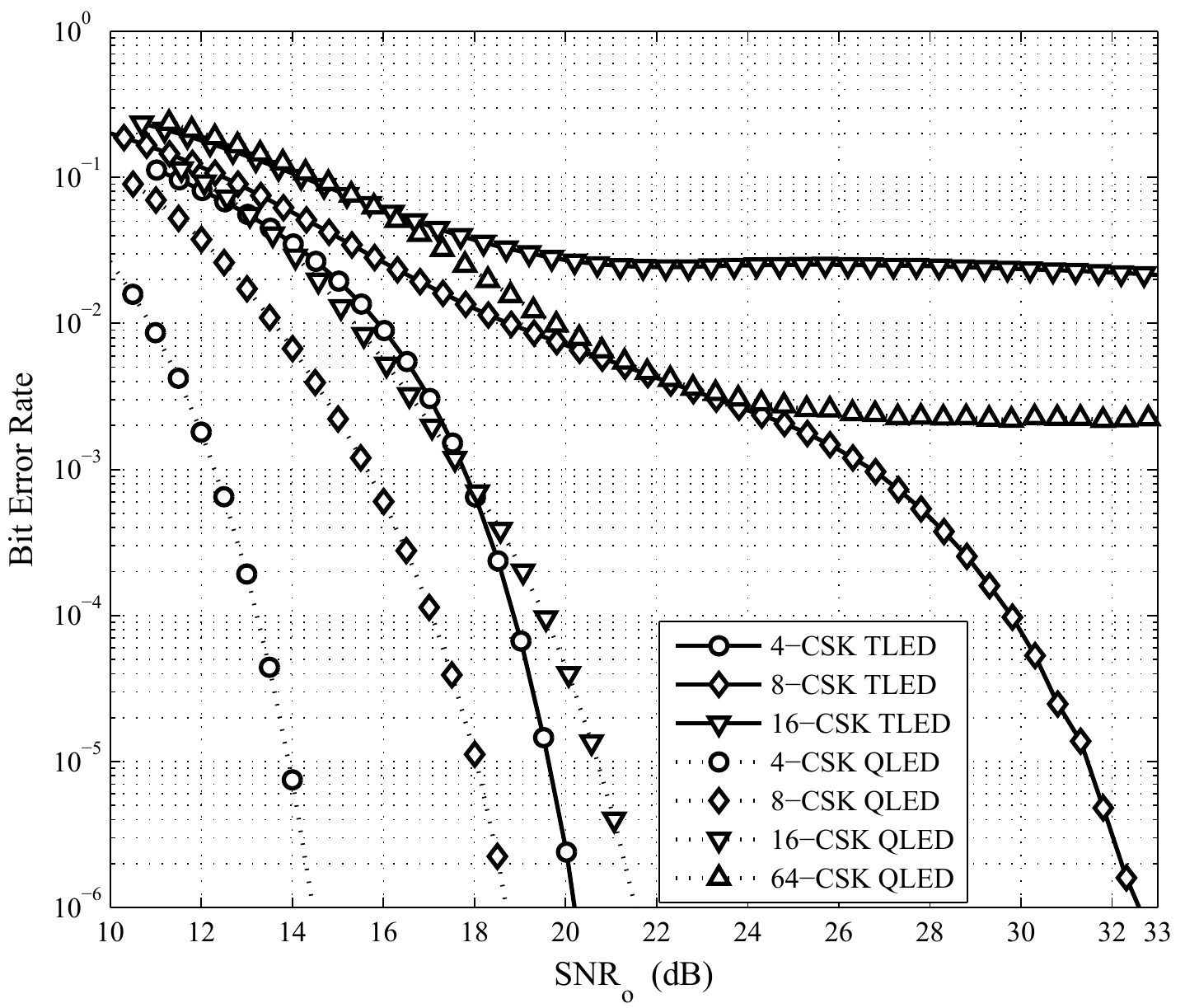}
	\caption{BER performance of uncoded and unequalised TLED and QLED system over optical wireless channel with $D_t=0.5$.}
	\label{fig:NoFDEberDt1}
\end{figure} 
\begin{figure}[tbph!]
	\centering
	\includegraphics[width=\linewidth]{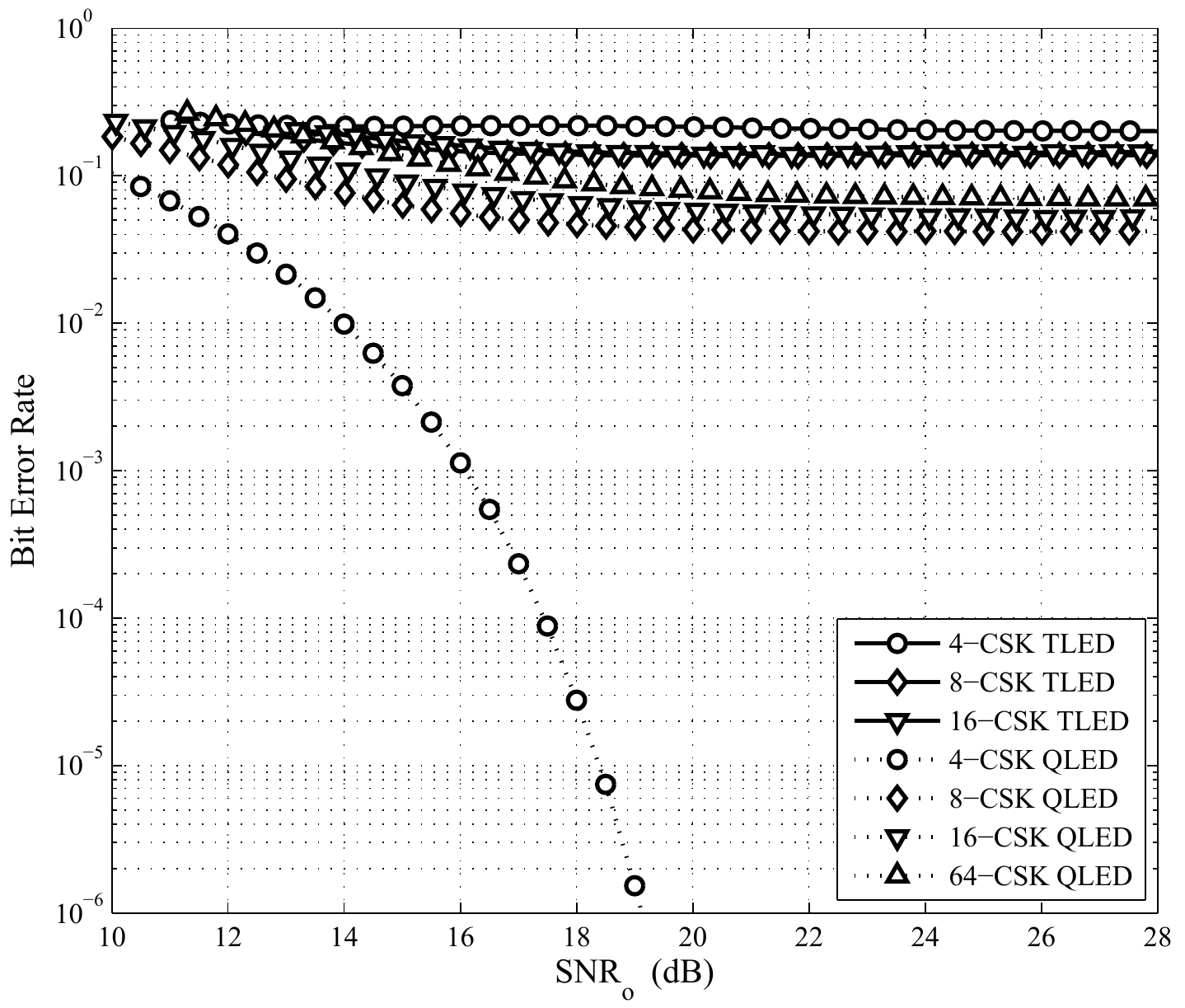}
	\caption{BER performance of uncoded and unequalised TLED and QLED system over optical wireless channel with $D_t=1$.}
	\label{fig:NoFDEberDt2}
\end{figure}
The effect of FDE on the performance of TLED and QLED CSK schemes was investigated based on a range of different values for the normalised delay spread $D_t$, which is generally given as, $D_t = D_{rms}/T_b$, where $T_b$ is the bit duration. A constant symbol rate, $R_s$, of 24 MSPS (Mega symbol per second) was assumed. This gives maximum unequalised data rates of up to 96 Mbit/s (64-CSK) for the TLED scheme and 288 Mbit/s (4096-CSK) for the QLED scheme. The data rates for different modulation orders of CSK can be calculated as:
\begin{equation}
\mbox{Data Rate} = \left(\frac{N}{N_{SB}}\right)R_{s}log_{2}M
\end{equation}

At first, the effect of ISI on the two CSK schemes was analysed by evaluating the unequalised system BER over two different dispersive optical channels. Fig.~\ref{fig:NoFDEberDt1} and Fig.~\ref{fig:NoFDEberDt2} show the BER versus optical signal-to-noise ratio (SNR\textsubscript{o}) performance of the TLED and QLED systems over an optical dispersive channel with $D_t=0.5$ and $D_t=1$, respectively. 

The results in Fig.~\ref{fig:NoFDEberDt1} show that the 16-CSK of TLED and 64-CSK\footnote{It is apparent that the QLED modulation orders for $M$ greater than 64 will also suffer from irreducible BER due to decreased Euclidean distances between the data symbols.} of QLED schemes will not be able to provide a good data link as the SNR requirements tends to infinity for a reasonable amount of BER, such as 10\textsuperscript{-6}. At the same time, we can see that under this channel condition, the TLED scheme can only provide data rates of up to 72 Mbit/s whereas the QLED scheme can only provide data rates of up to 96 Mbit/s. Therefore, both schemes will not be able to provide the maximum data rates which they are designed for. It should be noted that when $D_t=0.5$, the electrical rms delay spread of the channel is half of the bit duration. Similarly, the results in Fig.~\ref{fig:NoFDEberDt2} show that when $D_t$ is increased to 1, only the 4-CSK modulation of the QLED scheme can provide a good data link over this severely dispersive channel, limiting its data rate to just 48 Mbit/s. The results also show that the standardised TLED scheme will not be able to work in this case at all. These results show that the CSK systems require some form of equalisation in order to provide a working data link and sufficiently high data rates.

\begin{figure}[tbph!]
	\centering
	\includegraphics[width=\linewidth]{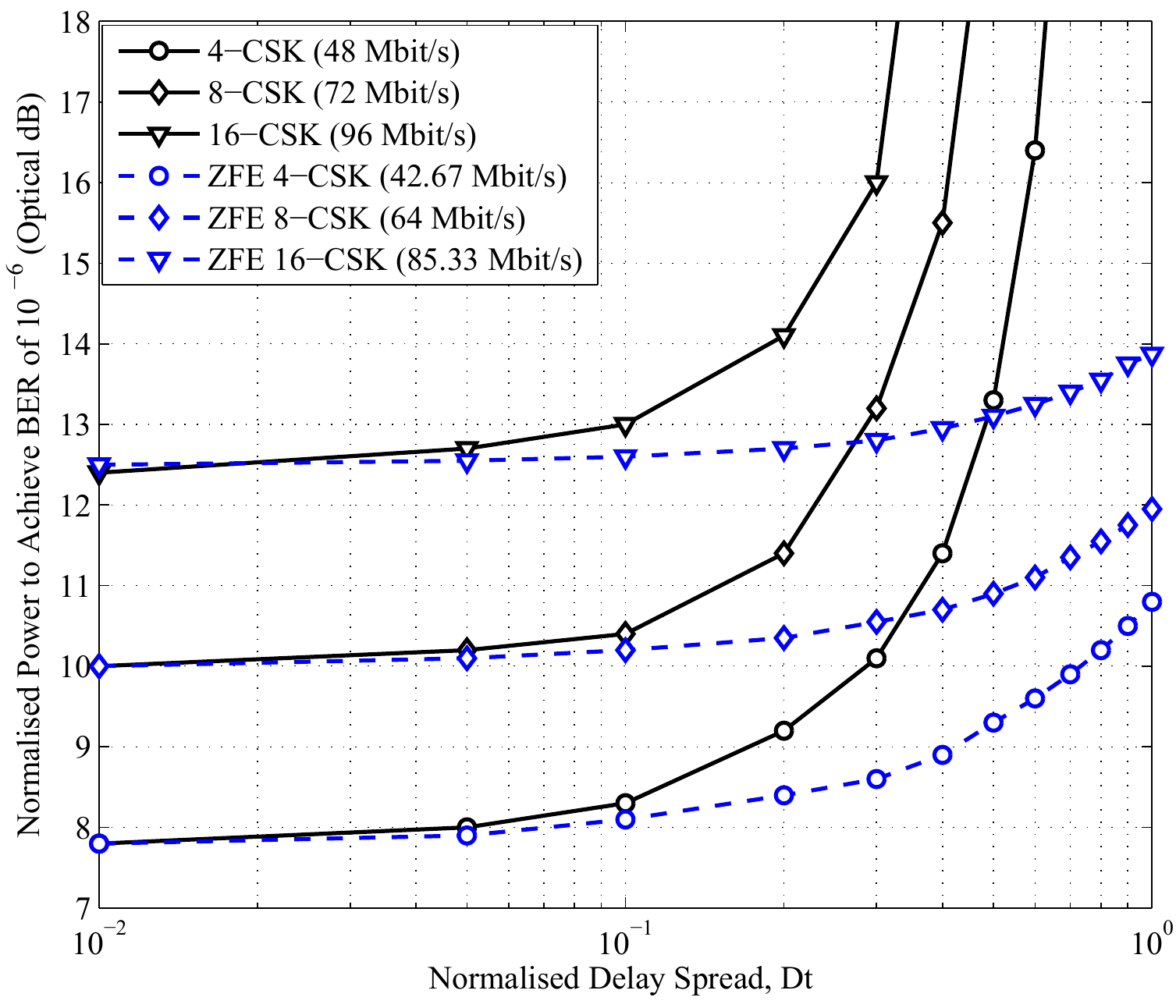}
	\caption{Dependence of unequalised and equalised multipath normalised power requirements on normalised delay spread, for TLED CSK modulations, to achieve a BER of 10\textsuperscript{-6}. All the power requirements are normalised relative to the optical power required by OOK over an AWGN channel.}
	\label{fig:TLEDzfeResults}
\end{figure}

\begin{figure}[tbph!]
\centering
\includegraphics[width=\linewidth]{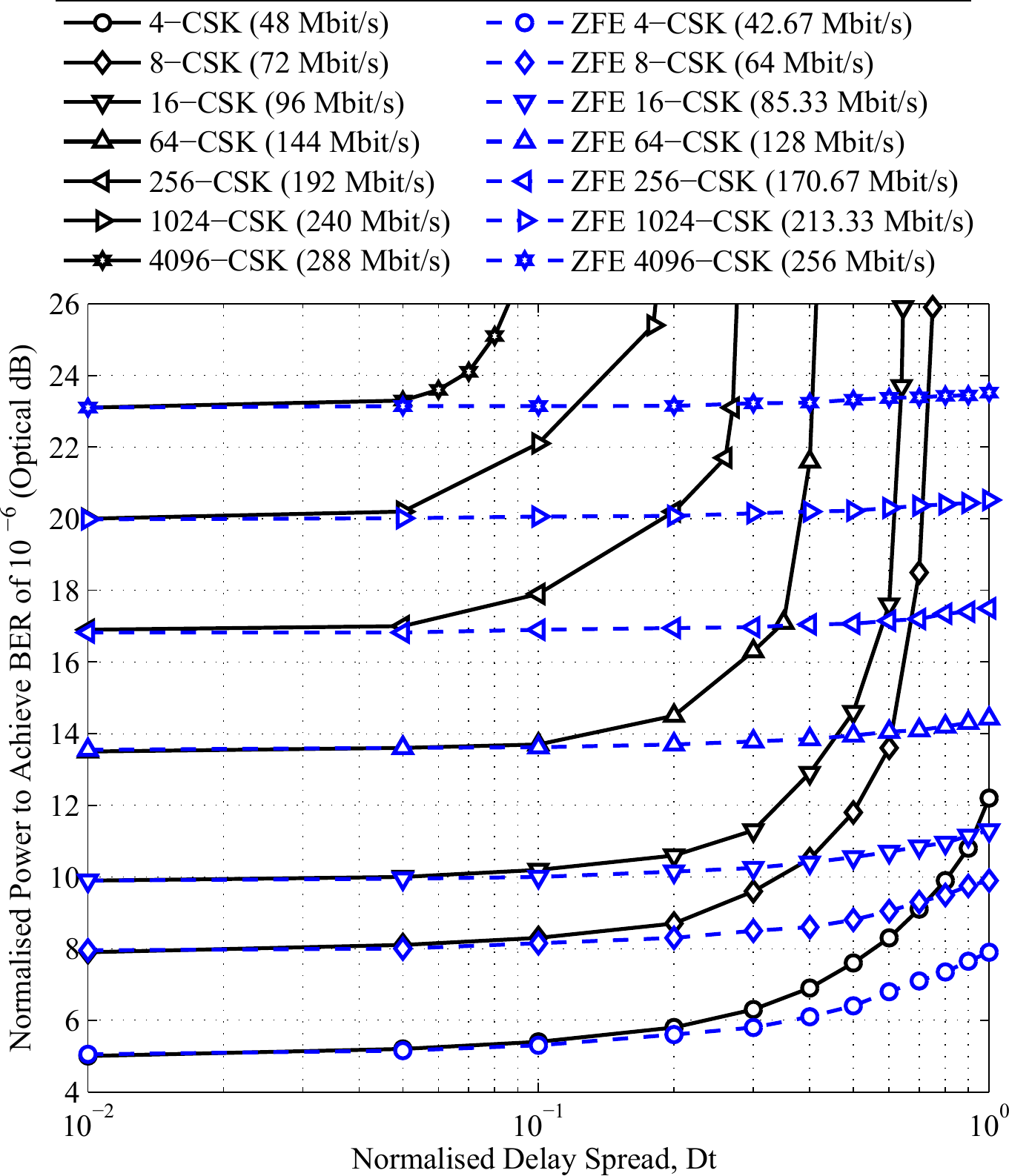}
\caption{Dependence of unequalised and equalised multipath normalised power requirements on normalised delay spread, for QLED CSK modulations, to achieve a BER of 10\textsuperscript{-6}. All the power requirements are normalised relative to the optical power required by OOK over an AWGN channel.}
\label{fig:QLEDzfeResults}
\end{figure}

Following this investigation, the performance of TLED and QLED CSK systems was evaluated over dispersive channels with $D_t$ ranging from 0.01 to 1, with and without the use of FDE and the results are presented in Fig.~\ref{fig:TLEDzfeResults} and Fig.~\ref{fig:QLEDzfeResults}, respectively. As mentioned previously, a zero-forcing algorithm based FDE was applied to both the TLED and QLED CSK schemes. During the simulations, $N$ of 64 was used to ensure that the sub-block bandwidth is much smaller than the smallest channel coherence bandwidth at $D_t$ of 1. The $L$ was set to 8 based on the maximum number of optical channel taps observed.

Fig.~\ref{fig:TLEDzfeResults} and Fig.~\ref{fig:QLEDzfeResults} show the optical power requirements of the uncoded-unequalised and uncoded-FDE based TLED and QLED CSK schemes, respectively, over a scale of $D_t$. The results show that the FDE (ZFE) based CSK modulation schemes offer large reductions in the optical power requirements, as $D_t$ increases, when compared to the unequalised CSK schemes. The FDE enables both QLED and TLED CSK to operate at their highest data rates using a finite amount of optical power over non-LOS optical channels with large delay spreads. Fig.~\ref{fig:TLEDzfeResults} and Fig.~\ref{fig:QLEDzfeResults} also show that the data rates of each modulation scheme reduce by $\sim$11.11\% when FDE is used due to the CP overhead. This gives a highest data rate of 85.33 Mbit/s for the TLED and 256 Mbit/s for the QLED scheme.

Using the results in Fig.~\ref{fig:TLEDzfeResults} and Fig.~\ref{fig:QLEDzfeResults}, Table~\ref{tab:QLEDvsTLEDinDis} compares the optical power requirements of both the CSK schemes, with and without the use of an equaliser, for three optical channels with different $D_t$. Table~\ref{tab:QLEDvsTLEDinDis} shows that FDE does not only allow higher data rate transmission for CSK schemes by enabling good data links for higher order modulations, but it also reduces the power requirements of the lower order modulation modes which require high optical powers without the use of FDE. The optical power reductions with FDE range from 0.08 dB to 12.6 dB\footnote{This optical power reduction is equivalent to approximately 27 dB in electrical domain.}, for any working modulation order in an unequalised case. However, theoretically it can be said that the power reductions are as high as $\infty$ dB.   

Table~\ref{tab:QLEDvsTLEDinDis} also shows that for the target BER of 10\textsuperscript{-6} the QLED CSK is more robust to channel dispersion effects than the TLED CSK. It can be seen that the 4-CSK of QLED scheme is the only modulation which can achieve the target BER without using FDE for a finite amount of optical power over a channel with $D_t=1$. Whereas with the use of FDE, the QLED scheme can provide the same data rates as the TLED scheme requiring 2.05 to 2.9 dB less optical power. This power gain of QLED scheme is mainly due to larger Euclidean distances between symbols restored after equalisation.  
\begin{table}[htbp!]
\renewcommand{\arraystretch}{1.2}
\centering
\caption{Normalised optical power requirements of uncoded-unequalised and uncoded-FDE based TLED and QLED CSK systems for a $D_t$ of 0.1, 0.5 and 1.}
\label{tab:QLEDvsTLEDinDis}
\begin{tabular}{|>{\centering\arraybackslash}p{0.1cm}|>{\centering\arraybackslash}p{0.2cm}|>{\centering\arraybackslash}p{1.3cm}|>{\centering\arraybackslash}p{1.57cm}|>{\centering\arraybackslash}p{1.57cm}|>{\centering\arraybackslash}p{1.57cm}|}
\hline \multicolumn{3}{|c|}{\raisebox{-20pt}{Modulation Schemes}} & Optical Power Requirements for $D_t=0.1$ (dB) & Optical Power Requirements for $D_t=0.5$ (dB) & Optical Power Requirements for $D_t=1$ (dB) \\ \hline
\hline
\multirow{10}{*}{\begin{turn}{90}
Unequalised
\end{turn}}& \multirow{3}{*}{\begin{turn}{90}
TLED
\end{turn}} & 4-CSK & 8.3 & 16.4 & $\infty$ \\ \cline{3-6}
& & 8-CSK & 10.4 & $\infty$  & $\infty$ \\ \cline{3-6}
& & 16-CSK & 13 & $\infty$  & $\infty$ \\ \cline{2-6}
&
\multirow{7}{*}{\begin{turn}{90}
QLED
\end{turn}}& 4-CSK & 5.4 & 7.6 & 20.5 \\ \cline{3-6}
 & & 8-CSK & 8.3 & 11.8 & $\infty$ \\ \cline{3-6}
 & & 16-CSK & 10.2 & 14.6 & $\infty$ \\ \cline{3-6}
 & & 64-CSK & 13.7 & $\infty$ & $\infty$ \\\cline{3-6}
 & & 256-CSK & 17.9 & $\infty$ & $\infty$ \\\cline{3-6}
 & & 1024-CSK & 22.1 & $\infty$ & $\infty$ \\\cline{3-6}
 & & 4096-CSK & $\infty$ & $\infty$ & $\infty$ \\
\hline
\hline
\multirow{10}{*}{\begin{turn}{90}
FDE-ZFE
\end{turn}}& \multirow{3}{*}{\begin{turn}{90}
TLED
\end{turn}} & 4-CSK & 8.1 & 9.3 & 10.8 \\ \cline{3-6}
& & 8-CSK & 10.2 & 10.9 & 11.95 \\ \cline{3-6}
& & 16-CSK & 12.6 & 13.1  & 13.87 \\ \cline{2-6}
&
\multirow{7}{*}{\begin{turn}{90}
QLED
\end{turn}}& 4-CSK & 5.3 & 6.4  & 7.9 \\ \cline{3-6}
 & & 8-CSK & 8.15 & 8.8 & 9.9 \\ \cline{3-6}
 & & 16-CSK & 10 & 10.55  & 11.3 \\ \cline{3-6}
 & & 64-CSK & 13.62 & 13.95 & 14.42 \\ \cline{3-6}
 & & 256-CSK & 16.9 & 17.07 & 17.5 \\ \cline{3-6}
 & & 1024-CSK & 20.06 & 20.22 & 20.52 \\ \cline{3-6}
 & & 4096-CSK & 23.14 & 23.32 & 23.52 \\
\hline
\end{tabular}

\end{table}
\section{Conclusion}
\label{sec:conclusion}
The use of low complexity FDE has been proposed for the standardised TLED and a recently designed QLED CSK schemes, while operating over a non-LOS dispersive optical channel. The performance analysis of the CSK schemes suggests that, without the use of FDE, the higher order modulation modes of both CSK systems will require infinite amount of optical power to operate over channels with moderate to high levels of dispersion, making it impossible for the CSK to provide a high data rate link. Though FDE will increase the overall system complexity in CSK, it will enable the higher order modulation modes to successfully operate even when the LOS is not available. Further, the complexity of FDE is significantly less than that associated with time domain equalisations for large delay spreads. The results show that the FDE will also substantially reduce the optical power requirements of CSK by 0.8 to 12.6 dB. For the considered delay spreads in the simulations, the FDE will add an overhead of up to 12.5\%.

\bibliographystyle{IEEEtran}
\bibliography{IEEEabrv,ref}

\begin{thebibliography}{10}
\providecommand{\url}[1]{#1}
\csname url@samestyle\endcsname
\providecommand{\newblock}{\relax}
\providecommand{\bibinfo}[2]{#2}
\providecommand{\BIBentrySTDinterwordspacing}{\spaceskip=0pt\relax}
\providecommand{\BIBentryALTinterwordstretchfactor}{4}
\providecommand{\BIBentryALTinterwordspacing}{\spaceskip=\fontdimen2\font plus
\BIBentryALTinterwordstretchfactor\fontdimen3\font minus
  \fontdimen4\font\relax}
\providecommand{\BIBforeignlanguage}[2]{{%
\expandafter\ifx\csname l@#1\endcsname\relax
\typeout{** WARNING: IEEEtran.bst: No hyphenation pattern has been}%
\typeout{** loaded for the language `#1'. Using the pattern for}%
\typeout{** the default language instead.}%
\else
\language=\csname l@#1\endcsname
\fi
#2}}
\providecommand{\BIBdecl}{\relax}
\BIBdecl

\bibitem{rajbhandari2017review}
S.~Rajbhandari, J.~J. McKendry, J.~Herrnsdorf, H.~Chun, G.~Faulkner, H.~Haas,
  I.~M. Watson, D.~O’Brien, and M.~D. Dawson, ``A review of gallium nitride
  leds for multi-gigabit-per-second visible light data communications,''
  \emph{Semiconductor Science and Technology}, vol.~32, no.~2, p. 023001, 2017.

\bibitem{zafar2017laser}
F.~Zafar, M.~Bakaul, and R.~Parthiban, ``Laser-diode-based visible light
  communication: Toward gigabit class communication,'' \emph{IEEE
  Communications Magazine}, vol.~55, no.~2, pp. 144--151, 2017.

\bibitem{IEEESTD}
``{IEEE Standard for Local and Metropolitan Area Networks--Part 15.7:
  Short-Range Wireless Optical Communication Using Visible Light},'' \emph{IEEE
  Std 802.15.7-2011}, pp. 1 --309, 6 2011.

\bibitem{PHYSummary}
R.~Roberts, S.~Rajagopal, and S.-K. Lim, ``Ieee 802.15.7 physical layer
  summary,'' in \emph{GLOBECOM Workshops (GC Wkshps), 2011 IEEE}, 2011, pp.
  772--776.

\bibitem{ourPaper}
R.~Singh, T.~O'Farrell, and J.~P. David, ``Performance evaluation of ieee
  802.15.7 csk physical layer,'' in \emph{Globecom Workshops (GC Wkshps), 2013
  IEEE}, Dec 2013, pp. 1064--1069.

\bibitem{sarbazi2013phy}
E.~Sarbazi and M.~Uysal, ``Phy layer performance evaluation of the ieee 802.15.
  7 visible light communication standard,'' in \emph{Optical Wireless
  Communications (IWOW), 2013 2nd International Workshop on}.\hskip 1em plus
  0.5em minus 0.4em\relax IEEE, 2013, pp. 35--39.

\bibitem{singh2015higher}
R.~Singh, T.~O'Farrell, and J.~P. David, ``Higher order colour shift keying
  modulation formats for visible light communications,'' in \emph{Vehicular
  Technology Conference (VTC Spring), 2015 IEEE 81st}.\hskip 1em plus 0.5em
  minus 0.4em\relax IEEE, 2015, pp. 1--5.

\bibitem{Drost}
R.~Drost and B.~Sadler, ``Constellation design for color-shift keying using
  billiards algorithms,'' in \emph{GLOBECOM Workshops (GC Wkshps), 2010 IEEE},
  2010, pp. 980--984.

\bibitem{EricSteve}
E.~Monteiro and S.~Hranilovic, ``Constellation design for color-shift keying
  using interior point methods,'' in \emph{Globecom Workshops (GC Wkshps), 2012
  IEEE}, 2012, pp. 1224--1228.

\bibitem{MM}
P.~Butala, J.~Chau, and T.~Little, ``Metameric modulation for diffuse visible
  light communications with constant ambient lighting,'' in \emph{Optical
  Wireless Communications (IWOW), 2012 International Workshop on}, oct. 2012,
  pp. 1 --3.

\bibitem{JLT}
R.~Singh, T.~OFarrell, and J.~David, ``An enhanced color shift keying
  modulation scheme for high-speed wireless visible light communications,''
  \emph{Lightwave Technology, Journal of}, vol.~32, no.~14, pp. 2582--2592,
  July 2014.

\bibitem{CIM}
K.-I. Ahn and J.~Kwon, ``Color intensity modulation for multicolored visible
  light communications,'' \emph{Photonics Technology Letters, IEEE}, vol.~24,
  no.~24, pp. 2254--2257, 2012.

\bibitem{singh2015physical}
R.~Singh, ``Physical layer techniques for indoor wireless visible light
  communications,'' Ph.D. dissertation, University of Sheffield, 2015.

\bibitem{singh2018coded}
R.~Singh, T.~O’Farrell, M.~Biagi, and J.~P. David, ``Coded color shift keying
  with frequency domain equalization for low complexity energy efficient indoor
  visible light communications,'' \emph{Physical Communication}, 2018.

\bibitem{singh2015analysis}
R.~Singh, T.~O'Farrell, and J.~P. David, ``Analysis of forward error correction
  schemes for colour shift keying modulation,'' in \emph{Personal, Indoor, and
  Mobile Radio Communications (PIMRC), 2015 IEEE 26th Annual International
  Symposium on}.\hskip 1em plus 0.5em minus 0.4em\relax IEEE, 2015, pp.
  575--579.

\bibitem{FDE1}
D.~Falconer, S.~Ariyavisitakul, A.~Benyamin-Seeyar, and B.~Eidson, ``Frequency
  domain equalization for single-carrier broadband wireless systems,''
  \emph{Communications Magazine, IEEE}, vol.~40, no.~4, pp. 58--66, Apr 2002.

\bibitem{oOFDM}
J.~Armstrong and B.~Schmidt, ``Comparison of asymmetrically clipped optical
  ofdm and dc-biased optical ofdm in awgn,'' \emph{Communications Letters,
  IEEE}, vol.~12, no.~5, pp. 343--345, May 2008.

\bibitem{Pridist1}
H.~Elgala, R.~Mesleh, and H.~Haas, ``Predistortion in optical wireless
  transmission using ofdm,'' in \emph{Hybrid Intelligent Systems, 2009. HIS
  '09. Ninth International Conference on}, vol.~2, Aug 2009, pp. 184--189.

\bibitem{Pridist2}
R.~Mesleh, H.~Elgala, and H.~Haas, ``An overview of indoor ofdm/dmt optical
  wireless communication systems,'' in \emph{Communication Systems Networks and
  Digital Signal Processing (CSNDSP), 2010 7th International Symposium on},
  July 2010, pp. 566--570.

\bibitem{CIE}
CIE, ``{Commission Internationale de l’Eclairage Proc.}'' 1931.

\bibitem{ZFE}
A.~Nuwanpriya, J.~Zhang, A.~Grant, S.-W. Ho, and L.~Luo, ``Single carrier
  frequency domain equalization based on on-off-keying for optical wireless
  communications,'' in \emph{Wireless Communications and Networking Conference
  (WCNC), 2013 IEEE}, April 2013, pp. 4272--4277.

\bibitem{WIC}
J.~Kahn and J.~Barry, ``Wireless infrared communications,'' \emph{Proceedings
  of the IEEE}, vol.~85, no.~2, pp. 265--298, feb 1997.

\bibitem{MoIRWC}
J.~Carruthers and J.~Kahn, ``Modeling of nondirected wireless infrared
  channels,'' \emph{Communications, IEEE Transactions on}, vol.~45, no.~10, pp.
  1260--1268, 1997.

\end{thebibliography}

\end{document}